\documentclass[aps,prl,floatfix,twocolumn,showpacs,superscriptaddress]{revtex4}
\usepackage{graphicx}
\usepackage{amsmath}
\usepackage{amssymb}
\usepackage{float}

\newcommand{\Emu}{E_{\mu}}
\newcommand{\Enu}{E_{\nu}}
\begin{document}

\title{Understanding Paramagnetic Spin Correlations \\in the Spin-Liquid
  Pyrochlore Tb$_2$Ti$_2$O$_7$}
\author{Ying-Jer Kao}
\email{y2kao@uwaterloo.ca}
\author{Matthew Enjalran}
\author{Adrian Del Maestro}
\author{Hamid R. Molavian }
\affiliation{ Department of Physics, University of Waterloo, Waterloo, ON,
   Canada N2L 3G1} 
\author{ Michel J. P.  Gingras}
\affiliation{ Department of Physics, University of Waterloo, Waterloo, ON,
   Canada N2L 3G1} 
\affiliation{ Canadian Institute for Advanced Research, Toronto, ON,
  Canada M5G 1Z8 }
\date{\today}
\begin{abstract}
Recent elastic and inelastic neutron scattering studies of the highly frustrated pyrochlore antiferromagnet
Tb$_2$Ti$_2$O$_7$ have shown some very intriguing features that cannot be
modeled by the local $\langle 111 \rangle$ classical Ising model, naively
expected to describe this system at low temperatures. Using the random phase
approximation to take into account fluctuations between the ground state doublet
and the first excited doublet, we successfully describe the elastic neutron
scattering pattern and dispersion relations in Tb$_2$Ti$_2$O$_7$, semi-quantitatively consistent with
experimental observations. 
\end{abstract}




\maketitle

The search for the physical realization of a spin liquid in three
dimensions has been a long quest for the condensed matter
community. Recently, geometrically frustrated magnetic systems have been the focus of intensive
experimental and theoretical studies because it is  believed that 
geometrical frustration can inhibit the formation of
long-range order, thus enabling the system to remain paramagnetic down to low temperatures.     
Among three-dimensional systems, the pyrochlore lattice of corner-sharing
tetrahedra 
 have been studied extensively.
It has been shown theoretically and numerically \cite{Villain1979, Reimers1992, Moessner1998}
that for classical Heisenberg spins with nearest-neighbor
antiferromagnetic (AF) exchange, there is no transition to
long-range magnetic order at finite temperature. This makes AF materials
based on the pyrochlore lattice excellent candidates to search for the
low-temperature spin liquid state. A number of experimental
studies on insulating pyrochlore materials have been carried
out. Interestingly, most materials either develop long-range N\'{e}el order,
such as Gd$_2$Ti$_2$O$_7$ \cite{Ramirez2002} and Er$_2$Ti$_2$O$_7$
\cite{Champion2001},
 or reveal spin-glass behavior,
such as Y$_2$Mo$_2$O$_7$ \cite{Gardner2001a}.
The ``spin ice'' materials,  Ho$_2$Ti$_2$O$_7$
\cite{Harris1997,Bramwell2001} and Dy$_2$Ti$_2$O$_7$ \cite{Ramirez1999,denHertog2000} 
 exhibit low-temperature thermodynamic properties reminiscent of Pauling's
 ``water ice model'' \cite{GingrasReview}. In these systems, an effective
 ferromagnetic interaction is frustrated due to the single-ion
 $\langle 111 \rangle$ Ising anisotropy \cite{Bramwell1998,GingrasReview}.
 The behavior of these systems can be quantitatively
described by the $\langle 111 \rangle$ Ising spin model with
nearest-neighbor exchange and long-range dipolar interactions \cite{denHertog2000,Bramwell2001,GingrasReview}.


 Tb$_2$Ti$_2$O$_7$ shows, however, very different and intriguing behavior.
 It is believed that Tb$_2$Ti$_2$O$_7$ belongs to the same family of
$\langle 111 \rangle$ Ising systems as Dy$_2$Ti$_2$O$_7$ and
 Ho$_2$Ti$_2$O$_7$  but with an effective nearest neighbor AF
interaction \cite{Gardner1999,denHertog2000,Gingras2000,Gingras2002,Gardner2001}. The same spin model
that very successfully described the spin-ice systems predicts it to have a
noncollinear N\'{e}el $\mathbf{Q}=0$ order, with all spins pointing into  or out
of each tetrahedron,  at about 1~K \cite{denHertog2000}. In dramatic contrast, 
Tb$_2$Ti$_2$O$_7$  remains a spin liquid, or
``cooperative paramagnet'', down to 70mK \cite{Gardner1999,Gardner2001}. 
In addition, recent paramagnetic neutron
scattering studies show that the scattering pattern for this material
is not consistent with a $\langle 111 \rangle$ Ising model, while a
Heisenberg spin model \cite{Gardner2001, Enjalran} or some level of
 relaxation away from the $\langle
111 \rangle$ Ising model \cite{Yasui2002} can better describe the observed neutron
scattering pattern. These results suggest that the restoration of spin isotropy in
the system, despite its expected Ising-like nature at low
temperature \cite{Gingras2000,Siddharthan1999,Rosenkranz2000,RosenkranzPrivate}, is
essential in understanding the paramagnetic spin correlations. 
Inelastic neutron scattering studies have also been performed
on this system and partial softening of the magnetic excitations at an
energy of about 20~K has been
observed \cite{Gardner1999,Gardner2001,Kanada1999}. This has been 
attributed to a (spin) roton-like mode, as in liquid $^4$He \cite{Gardner1999,Gardner2001,Feynman}, which
further indicates a more isotropic  nature of the spins.
Given the ensemble of evidences, it would appear that one needs a more isotropic
spin model to understand the paramagnetic spin correlations in this
system. More importantly, such a ``restoration'' of spin isotropy may also be the key to 
understanding  why Tb$_2$Ti$_2$O$_7$ fails to order down to low temperatures. 
In this paper, we employ
the random-phase approximation (RPA) \cite{Jensen} to take into account the
fluctuations between the ground state doublet and the first excited 
doublet. We successfully describe the observed
paramagnetic spin correlations in Tb$_2$Ti$_2$O$_7$, without any assumptions
regarding the nature of
the spins, while we still obtain the $\mathbf{Q}=0$ N\'{e}el order at low temperatures. 
This result makes the fact that Tb$_2$Ti$_2$O$_7$ fails to order
even more puzzling.


We begin with the model spin Hamiltonian,
\begin{equation}
\mathcal{H}=\frac{1}{2}\sum_{i,j,\alpha,\beta,a,b} S^{\alpha}_{i,a}
\mathcal{J}^{\alpha\beta}_{ab}(i,j)
S^{\beta}_{j,b}+\sum_{i,a}\mathcal{H}_{\rm CF}(i,a),
\end{equation}
where  $i,j$ are  indices of the Bravais lattice vectors for the FCC lattice, 
$a,b$ are  indices of sublattice basis vectors and $\alpha,\beta$ are
indices for the spatial
coordinates. $\mathcal{H}_{\rm CF}$ is the single-ion crystal
field (CF) Hamiltonian. 
The spin-spin interaction matrix $\mathcal{J}$, including both exchange and
long-range dipolar interactions, reads
\begin{equation*}
\mathcal{J}^{\alpha\beta}_{ab}(i,j)=-J_1\delta^{\alpha\beta}\delta_{nn}+D_{dd}\left[
    \frac{\delta^{\alpha\beta}}{ |\mathbf{R}^{ab}_{ij}|^3}-\frac{3 R^{ab,\alpha}_{ij}
    R^{ab,\beta}_{ij}}{ |\mathbf{R}^{ab}_{ij}|^5}
\right], 
\end{equation*} 
where $\delta_{nn}$ refers to nearest-neighbor interaction, and
 $R^{ab,\alpha}_{ij}$ denotes
 the  $\alpha$ component of the interspin
vector $\mathbf{R}^{ab}_{ij}$ that connects spins $\mathbf{S}_{i,a}$ and
$\mathbf{S}_{j,b}$. 
The nearest-neighbor exchange for Tb$_2$Ti$_2$O$_7$ is given by
$J_1\approx-0.167$~K \cite{exchange}, and 
$D_{dd}=(g\mu_{B})^2 \mu_0/4\pi$, where $\mu_B$ is the Bohr magneton, $\mu_0$ the magnetic permeability, and
$g=3/2$ for Tb$^{3+}$. 

\begin{figure}
\includegraphics[width=3.2in,clip]{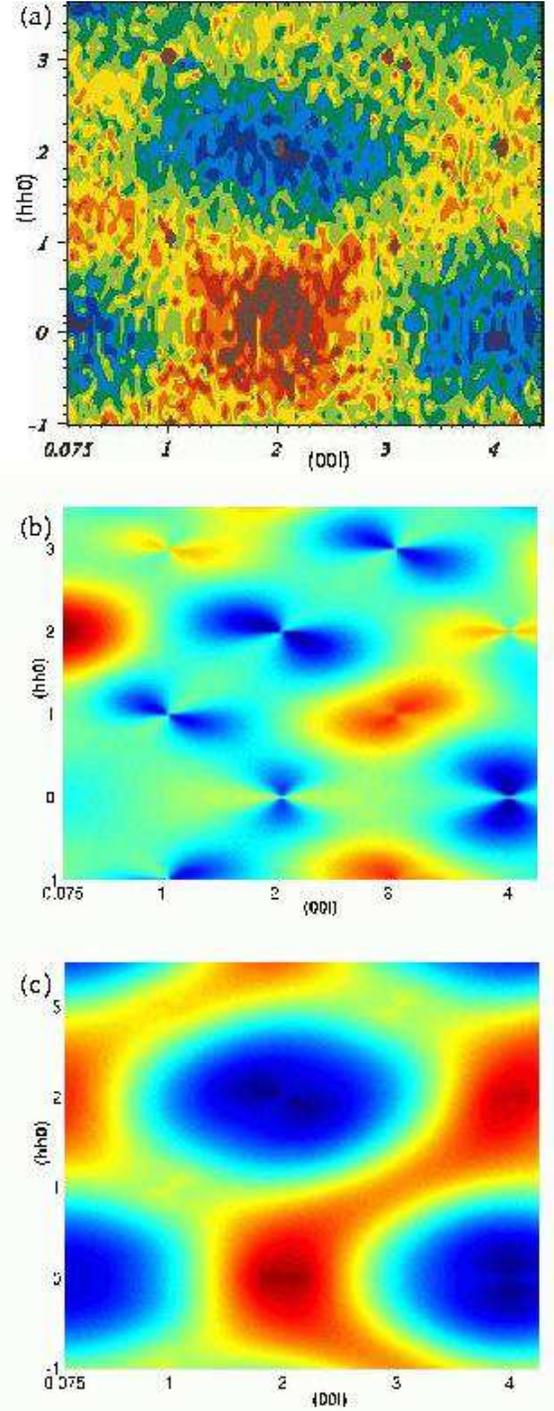}
\caption{\label{fig:scattering} 
(a) Experimental elastic neutron scattering pattern of Tb$_2$Ti$_2$O$_7$ in
  the $(h,h,l)$ plane of reciprocal space at $T$=9~K, from
  Ref. \onlinecite{Gardner2001}. Dark blue shows the lowest intensity level,
  red-brown the highest. 
  (b)  $S(\mathbf{Q})$ for the $\langle 111
  \rangle$ Ising model at $T$=9~K. (c) $S(\mathbf{Q})$ for the
  doublet-doublet model at $T$=9~K. Note that magnetic form factor $|F(\mathbf{Q})|^2$ is
  divided out in both experimental and theoretical results. }
\end{figure}

 The single-ion susceptibility is given by \cite{Jensen} 
\begin{eqnarray}
\chi^{0,\alpha\beta}_{a}(\omega)&=&\sum_{\mu,\nu}^{\Emu\ne\Enu}\frac{M^\alpha_{\nu\mu,a}
  M^\beta_{\mu\nu,a}}{\Emu-\Enu-\hbar(\omega+i 0^+)}(n_{\nu}-n_{\mu})\nonumber\\
& &+\;\frac{\delta(\omega)}{k_B T}\sum_{\mu,\nu}^{\Emu=\Enu}
M^\alpha_{\nu\mu,a}M^\beta_{\mu\nu,a}n_{\nu},
\label{eq:chi0}
\end{eqnarray}
where $n_{\nu}$ is the thermal occupation fraction for state $\nu$. 
The matrix elements for the single-ion states are given by 
$M^\alpha_{\nu\mu,a}=\sum_{\bar{\alpha}}\langle \nu| {S}^{\bar{\alpha}} | \mu
\rangle u^{\alpha}_{\bar{\alpha},a}$, 
where $u^{\alpha}_{\bar{\alpha},a}$ is the rotation matrix from the local
($\bar{\alpha}$) frame defined on sublattice $a$ to the global ($\alpha$) frame. 
The spin operator
${S}^{\bar{\alpha}}$ acts on the CF states  defined
along the local quantization axis \cite{Gingras2000}. 
The local wave function structure has been obtained from both experimental
measurements and theoretical calculations \cite{Gingras2000,Rosenkranz2000,RosenkranzPrivate}.
The primary components are $J_z$ eigenstates $|\pm 4 \rangle$ for the ground
state doublet and $|\pm5\rangle$ for the first-excited state doublet. The
doublets are 
separated by an anisotropy gap $\Delta\approx20$~K, which is comparable with the Curie-Weiss temperature
$\theta_{\rm CW}\approx-19$~K \cite{Gingras2000}. 

 The RPA equation which takes into account the two-ion interaction
 contribution to  the full susceptibility \cite{Jensen}
is given by
\begin{equation*}
\chi^{\alpha\beta}_{ab}(\mathbf{q},\omega)+\sum_{\gamma,\delta,c}
\chi^{0,\alpha\gamma}_{a}(\omega)
\mathcal{J}^{\gamma\delta}_{ac}(\mathbf{q})\chi^{\delta\beta}_{cb}(\mathbf{q},\omega)=
\delta^{\ }_{ab}\chi^{0,\alpha\beta}_{a}(\omega),
\end{equation*} 
where $\mathcal{J}(\mathbf{q})$ is the Fourier transformation of the
interaction matrix $\mathcal{J}(i,j)$. 
The slowly converging, infinite lattice sum of the dipolar
interaction is handled using Ewald summation
techniques \cite{Ewald,Enjalran}.   We solve for
$\chi^{\alpha\beta}_{ab}(\mathbf{q},\omega)$ numerically using LAPACK
routines. 


The elastic and dynamical neutron cross-sections are related to the spin
susceptibilities by summing over the sublattice
contributions and taking the transverse components of $\mathbf{S}$
perpendicular to $\mathbf{Q}$ \cite{Jensen},
\begin{subequations}
\begin{eqnarray}
S_{\rm el}(\mathbf{Q}) &\propto& \frac{|F(\mathbf{Q})|^2}{k_B
  T} \sum_{\alpha,\beta;a,b}(\delta_{\alpha\beta}-\hat{Q}_\alpha\hat{Q}_\beta)
  \nonumber\\
&\times &\exp\left[-i(\mathbf{r}^a-\mathbf{r}^b)\cdot
  \mathbf{G}\right]\mbox{Re}\,\chi_{ab}^{\alpha\beta}(\mathbf{q});\\
S_{\rm dyn}(\mathbf{Q},\omega) &\propto& \frac{|F(\mathbf{Q})|^2}{1-\exp(-\hbar
  \omega/k_B  T)} \sum_{\alpha,\beta;a,b}(\delta_{\alpha\beta}-\hat{Q}_\alpha\hat{Q}_\beta)
\nonumber  \\
&\times &\exp\left[-i(\mathbf{r}^a-\mathbf{r}^b)\cdot
  \mathbf{G}\right]\mbox{Im}\,\chi_{ab}^{\alpha\beta}(\mathbf{q},\omega),
\end{eqnarray}
\end{subequations}
where $\mathbf{Q=G+q}$,  $\mathbf{G}$ is a reciprocal lattice vector
of the FCC lattice, and $\mathbf{q}$ is a primitive vector in the first
Brillouin zone. $\mathbf{r}^a$ is the sublattice basis vector and $F(\mathbf{Q})$
is the magnetic form factor for Tb$^{3+}$.

Figure~\ref{fig:scattering}a shows the experimental scattering pattern in the
$(h,h,l)$ plane at 9~K \cite{Gardner2001}. Note that there is a strong intensity maximum around $(0,0,2)$.
Figure~\ref{fig:scattering}b shows the  $S(\mathbf{Q})$  calculation using the
$\langle 111 \rangle$ Ising model, i.e., the anisotropy gap
$\Delta\rightarrow\infty$. It is clear that this model fails to reproduce the
correct neutron scattering pattern as observed experimentally, as there is
only weak non-critical intensity around $(0,0,2)$. 
It has been shown that in the $\langle 111 \rangle$ Ising model,
the intensity at $(0,0,2)$ and $(0,0,0)$ are exactly correlated, and vanishes
for Tb$_2$Ti$_2$O$_7$ \cite{Enjalran,Gardner2001,Gingras2000}.
Using a more realistic
doublet-doublet model for Tb$_2$Ti$_2$O$_7$, we are able 
to qualitatively reproduce the experimentally observed scattering pattern (Fig.~\ref{fig:scattering}c).  
It captures most details of the experimental pattern, such as the intensity maximum
around $(0,0,2)$, the minima around $(0,0,0)$, $(2,2,2)$ and $(0,0,4)$. This excellent 
agreement between theory and experiment indicates that to properly understand
the spin correlations in Tb$_2$Ti$_2$O$_7$, the fluctuations out of the
ground state doublet to the first excited doublet, originating
from the first term in Eq.~\eqref{eq:chi0}, are important.
 It should be noted that this result is obtained from the
simple level scheme described above with no other assumptions about the nature of
the spin or the details of the wave functions.

\begin{figure}
\includegraphics[width=2.8in,clip]{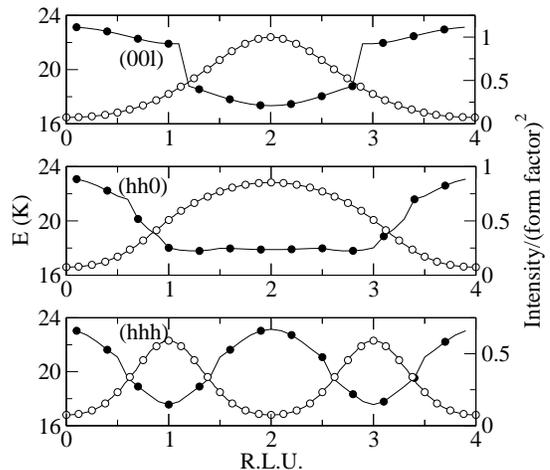}
\caption{\label{fig:inelastic}
The open symbols show the line scans of the $S(\mathbf{Q})$ for the
  three high-symmetry directions: from top to bottom, $(0,0,l)$, $(h,h,0)$ and
  $(h,h,h)$. The filled symbols show the dispersion relation for the
  lowest-lying branch of magnetic excitations at $T=4$~K. }
\end{figure}

Line scans in reciprocal space along three high-symmetry directions
$(0,0,l)$, $(h,h,0)$ and $(h,h,h)$  at $T=$4~K are plotted in
Fig.~\ref{fig:inelastic}.
 Open symbols are the data points for  $S(\mathbf{Q})$. 
Filled symbols  are the dispersion
relations $E(\mathbf{Q})$ for the lowest-energy band of magnetic excitations
at 4~K. 
The dispersion relations show good semi-quantitative
agreement with the experimental dispersion observed in single crystal Tb$_2$Ti$_2$O$_7$
reported in Fig.~9 of  Ref.~\onlinecite{Gardner2001}. We also find that the minimum in
$E(\mathbf{Q})$ corresponds to the maximum in $S(\mathbf{Q})$ in all
symmetry directions, as is observed in the experiment. The sharp
jumps in dispersions, e.g., near $(0,0,1)$ and $(0,0,3)$, correspond to  shifts of spectral
maximum between different branches, since the experiment and current calculation track
only the maximum peaks in the energy spectra. In principle,
there should be four branches of  magnetic
excitations with different intensities due to the four sublattice structure. Higher
resolution experiments need to be performed to map out the dispersion relations of these branches.     

\begin{figure}
\includegraphics[width=2.8in,clip]{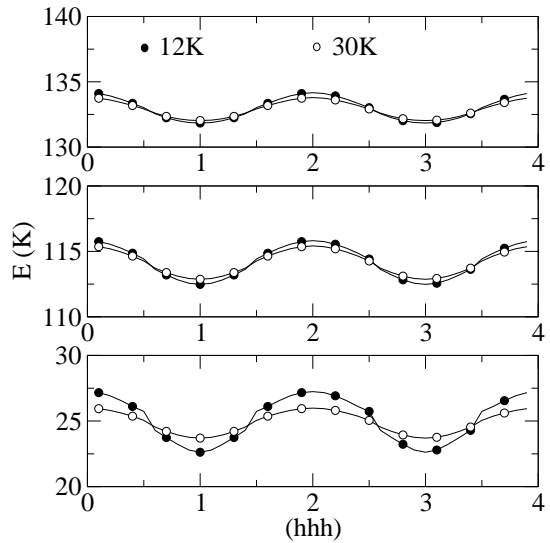}
\caption{\label{fig:temp}
The dispersion relations of the three lowest-lying magnetic
  excitations along the $(h,h,h)$ direction at both 12~K and 30~K. The lowest
  band displays a more pronounced energy dispersion as temperature is
  lowered, in agreement with experimental observations \protect\cite{Gardner2001}. }
\end{figure}

To study the temperature effects on the dispersion relations of the lowest
three bands of magnetic excitations, we use the wave functions obtained from
crystal field calculations in the point charge
approximation  and in the fixed $J=6$ manifold for
Tb$^{3+}$ \cite{Gingras2000}. This gives the CF level scheme with two
low-lying doublets as described above, and two higher-energy
singlets \cite{Gingras2000,Gardner2001}. The energy levels are given as 0,
24.6, 113.9 and 132.7K, consistent with experiments.  The two higher
singlets have large $|\pm3\rangle$ contributions
and some mixing with other  $|J,M_J\rangle$ components \cite{Gingras2000}.
 All the excited states are connected to
the ground state doublet through the $S^+$ and $S^-$ operations and the excitations are visible via neutron
spectroscopy \cite{Gardner2001}. Figure~\ref{fig:temp} shows the dispersion relations of the
three lowest-lying magnetic excitations along $(h,h,h)$ at 12~K and 30~K at the
same energy scale. The lowest band becomes more dispersive as the
temperature is lowered, while the higher two bands do not show much change
with temperature.


Fluctuations between  the doublets are important in understanding the 
experimentally observed  scattering pattern. These fluctuations
originate in  the first term of Eq.~\eqref{eq:chi0}, where the
anisotropy gap $\Delta=E_1-E_0$ enters the formulation algebraically, instead
of exponentially as in the elastic (second) term through the thermal
occupation fraction. If the two-ion
interaction, characterized by $\theta_{\rm CW}$, is comparable with
$\Delta$, then the
effect of this term is large and part
of the isotropic response is restored. 
By setting $\Delta\rightarrow\infty$ or the matrix element
between two doublets to zero, this term is eliminated, and  the scattering
pattern is reduced to that of the $\langle 111 \rangle $ Ising model
\cite{Enjalran}.
 This explains why the
$\langle 111 \rangle$ Ising model works so well in describing the spin
correlations in spin ice materials \cite{denHertog2000,Bramwell2001}, 
since in these systems  $\Delta \approx 200\sim300$~K \cite{Rosenkranz2000,RosenkranzPrivate} and
$\theta_{\rm CW}\approx 0.5\sim2$~K
\cite{denHertog2000,Bramwell2001,Ramirez1999}, so the effects of the fluctuations out of
the ground state doublet are negligible. To understand the different temperature dependencies
of the dispersion for the three lowest-lying magnetic excitations, we inspect the thermal
factor $n_{\nu\mu}=n_\nu-n_\mu$ in the first term of Eq.~\eqref{eq:chi0}. 
For temperatures much lower than $\Delta$, only
the ground state doublet is populated, and this factor is almost the same for
all three bands. When the temperature reaches a value
$T \approx \Delta$, the first excited state doublet begins to be populated, therefore
reducing the factor $n_{01}=n_0-n_1$, while those for the two
 higher levels remain essentially unchanged. This change in thermal occupation factor
results in a large temperature dependence of the dispersion in the lowest band. 


The low temperature ground state of our model in RPA is the 
noncollinear  ``all-in'' or ``all-out'' $\mathbf{Q}=0$  N\'{e}el state
\cite{denHertog2000} with $T_c\approx 1.8~$K. Exact $T_c$ value
depends on the details of the crystal field wave functions. Suprisingly,
this value is very close to the 
$T_c$ obtained recently in hydrostatic pressure measurements for
pressure greater than  1~GPa \cite{Mirebeau2002}. By standard Holstein-Primakoff
expansion aroud this ground state
we obtain the spin-wave excitation spectrum with a large gap  
and the reduction of the staggered magnetization due to quantum fluctuations
$\Delta m/m < 10^{-4}$\cite{Adrian}.    
This result should be taken only as an indicator, though,  
since the ground state doublet ($M_J=\pm4$) is not at its saturation value
$M_J=\pm6$ due to the crystal field interaction. The magnetic ground state doublet is
a non-Kramers doublet, and the states 
are time-conjugate of each other, so there exist no matrix elements between
them \cite{Abragam},
indicating that any quantum fluctuations
causing spins to flip result from  a higher-order virtual process via
excited crystal field levels.

In conclusion, with a simple level scheme and the RPA to include the fluctuations between the ground
state and the first excited state  doublets, we are able to
describe semi-quantitatively the experimentally observed neutron scattering pattern and the energy
dispersion in Tb$_2$Ti$_2$O$_7$. Our results indicate that the crystal field
effects are important in  this system
due to the fact that the anisotropy gap and the two-ion interaction are
comparable, and that some isotropy is present in the response. 
The ground state of our model, however, is still the ``all-in'' or
 ``all-out'' $\mathbf{Q}=0$ N\'{e}el state \cite{denHertog2000} with $T_c\approx 1.8~$K, which
 is stable against the lowest order quantum fluctuations.  
These results indicate that a mechanism that may restore more isotropy and thus increase the
quantum fluctuations leading to the suppression of $T_c$, requires further
detailed study.
 
We thank B. Buyers, B. Canals, B.~Gaulin,
J.~Gardner and  S.~Rosenkranz for useful and stimulating discussions. 
This work is supported by the NSERC of Canada, Research Corporation and the
Province of Ontario.


\begin{thebibliography}{29}
\bibitem[{exc()}]{exchange}
\bibinfo{note}{$J_1$ is estimated from $\theta^{\rm\,ex}_{\rm CW}\approx
  -14$~K, where the crystal field contribution is subtracted off from an
  effective paramanetic $\theta_{\rm CW}\approx -19$~K
  (Ref.~\onlinecite{Gingras2000}), and $J=6$ for Tb$^{3+}$.}
\end{thebibliography}

\end{document}